\documentclass{aa}
\usepackage[varg]{txfonts}
\bibpunct{(}{)}{;}{a}{}{,} 
\usepackage{multirow}
\usepackage{amsmath}
\usepackage{mathtools}

\newcommand{\hii}{\ion{H}{ii}}
\newcommand{\hh}{\ion{H}{ii}~}

\newcommand{\nii}{[\ion{N}{ii}]}

\newcommand{\oii}{[\ion{O}{ii}]}
\newcommand{\oiii}{[\ion{O}{iii}]}
\newcommand{\sii}{[\ion{S}{ii}]}
\newcommand{\siii}{[\ion{S}{iii}]}

\begin{document}

\title{Inner and outer star forming regions over the disks of spiral galaxies} 

\subtitle{II. A comparative of physical properties and evolutionary stages} 

\author{M. Rodr\'{i}guez-Baras\inst{1} \and A.I. D\'{i}az\inst{1} \and F.F. Rosales-Ortega\inst{2}} 


\institute{Departamento de F\'{i}sica Te\'orica, Universidad Aut\'onoma de Madrid, 28049 Madrid, Spain.
  \and Instituto Nacional de Astrof{\'i}sica, {\'O}ptica y Electr{\'o}nica, Luis E. Erro 1, 72840 Tonantzintla, Puebla, M\'exico.}

\date{\today}

\abstract 
{The \hh regions are all studied employing the same general prescriptions and models, independently of the regions location in the galaxy disk. However, observed discrepancies between physical properties of inner and outer regions may indicate systematic differences in their star formation processes due to the influence of their environments.}
{Through the analysis of inner and outer \hh region observed spectra, we aim to explore possible systematic differences between the physical properties (metallicity, mass, and age) of their ionising clusters in order to study how star formation proceeds in different environments.}
{We analysed two samples of 725 inner and 671 outer regions, characterised in the first paper of this series. Their functional parameters (oxygen abundances, ionisation parameters, and effective temperatures) were estimated and this parameter grid is employed as input for the computation of 540 Cloudy photoionisation models. Observed regions are compared with model predictions using diagnostic and evolutionary diagrams. }
{Higher metallicities are confirmed for the inner regions, although there are important discrepancies between the diagnostic diagrams. Calibrations based on the N2 index may underestimate inner regions oxygen abundances due to the \nii~saturation at solar metallicities. The degeneracy between the age and ionisation parameter affects oxygen abundance calibrations based on the O3N2 index. Innermost regions seem to have enhanced N/O ratios with respect to the expected values considering secondary production of nitrogen, which indicate an increase in the slope of the relation between N/O and O/H. Ionisation parameter calibrations based on the  \sii/H$\alpha$ ratio are not valid for inner regions due to the observed bivalued behaviour of this ratio with O/H. Innermost regions have lower \oii/\oiii~ratio values than expected, indicating a possible non-linear relation between \textit{u} and \textit{Z}. Composite stellar populations (ionising and non-ionising) are present in both inner and outer regions, with an ionising contribution of around 1\%. In considering the effects of evolution and underlying populations, inner regions show larger ionising cluster masses that possibly compose star-forming complexes. The most conservative lower limit for ionising cluster masses of outer regions indicate that they might be affected by stochastic effects. Equivalent widths indicate younger ages for outer regions, but degeneracy between evolution and underlying population effects prevent a quantitative determination. Nebular properties of the \hh regions are also derived: inner regions have larger angular sizes, lower filling factors, and larger ionised hydrogen masses.}
{Systematic physical differences are confirmed between ionising clusters of inner and outer \hh regions. These differences condition the validity and range of reliability of oxygen abundance and ionisation parameter calibrations commonly applied to the study of \hh regions.}

\maketitle

\section{Introduction}
\label{Sec: Introduction}

By definition, the gas ionisation in \hh regions is due to recently formed high mass stars, ranging from both single stars to star clusters. In the latter case, their large luminosities make them easily observable on the disks of spiral galaxies. Setting HST data or the use of adaptive optics techniques aside, the embedded ionising clusters cannot be resolved, except in nearby galaxies (e.g. 30Dor in the LMC, \cite{1997ApJS..112..457W}; NGC604 in M33, \cite{1996ApJ...456..174H}; NGC5471 in M101, \cite{2011AJ....141..126G}). However, their properties can be inferred from integrated observations of the ionised nebulae. Therefore, observing \hh regions that are very close to the centres of galaxies and to the outskirts of their disks provides the means to explore how star formation proceeds in different environments.

The emission line spectra of \hh regions, independent of their location in a galaxy, look rather similar since they respond to the local conditions of the gas. Regarding this, the most relevant agent in conforming the emission line ratios of different elements present in the spectra is probably the metal content as it controls the cooling of the ionised nebula and, therefore, its electron temperature. Nevertheless, other properties like the radiation field to which they are exposed, the gas density, and geometrical factors should not be disregarded. Conversely, it is possible to derive the elemental abundances of the region by measuring its observed emission line ratios from a given spectrum. For example, oxygen abundance radial gradients in spiral galaxies can be obtained from \hh region observations \citep[e.g.][]{1988MNRAS.235..633V,2003ApJ...591..801K} and are essential for testing galaxy formation scenarios \citep{2014A&A...563A..49S} and chemical evolution models \citep[see][]{2005MNRAS.358..521M,2017MNRAS.468..305M}. 
 
Even recognising that, in doing so, the same prescriptions and models are used in all cases implicitly, the higher abundances found close to the central regions of galaxies will affect stellar evolution through the increased opacity of the stellar material, so that high metallicity stars will have lower effective temperatures. This should be reflected in an inverse correlation between the metallicity and temperature of the ionising radiation of \hh regions, with the ionising stars having a lower effective temperature in inner, higher metallicity, regions than in outer, lower metallicity, ones. 

Lower effective temperatures imply lower photon fluxes. On the other hand, according to stellar evolution models \citep{1991A&A...242...93M,2001A&A...369..574V}, a higher metallicity implies a larger number of the Wolf-Rayet stars (WR) and, therefore, a higher amount of mechanical energy provided by their winds. Both effects taken together would lead to lower ionisation parameters for inner \hh regions versus outer ones \citep{2006ApJS..167..177D,2009MNRAS.398..451M}. The proof of this is still controversial; \cite{1996ApJ...456..504K} found a weak correlation, while \cite{1999ApJ...510..104B} found a strong one. More recently, \cite{2011MNRAS.415.3616D}, from the application of photoionisation models to a large \hh region compiled sample, find no correlation between the ionisation parameter and metallicity.

This is the second article of a series of three, and it is focused on the comparative analysis and study of the physical properties of the inner and outer \hh region observational samples previously characterised in \cite{2018A&A...609A.102R}, hereafter Paper I. The CALIFA survey provides the opportunity of analysing the kind of imaging and integrated spectroscopy required to extract the quantitative information needed to characterise the physical properties of the ionising clusters. At the same time, the homogeneity of data and their treatment is guaranteed. Section 2 provides a summary of the observations, data reduction and the selection criterion for \hh region extraction, already presented in Paper I. Section 3 is devoted to the estimation of the \hh regions functional parameters. Section 4 describes the photoionisation models employed to interpret the data through the use of diagnostic diagrams. Section 5 presents the derived quantities of the inner and outer \hh region samples. Section 6 describes the evolutionary stages of the ionising populations. Finally, our summary and conclusions are given in Sect. 7.

\section{Properties of the observed \hh regions}
\label{Sec: Properties of the observed regions}

\subsection{Observations and data reduction}
\label{Subsec: Observations and data reduction}

The \hh regions comprising this work belong to a sample of galaxies part of the CALIFA survey \citep{2012A&A...538A...8S}. CALIFA observations were carried out at the Centro Astron\'{o}mico Hispano-Alem\'{a}n (CAHA) 3.5m telescope using the Potsdam Multi Aperture spectrograph \citep[PMAS; ][]{2005PASP..117..620R} in the PPAK mode \citep{2004AN....325..151V,2006NewAR..50..355K,2006PASP..118..129K}, a retrofitted bare fibre bundle IFU which expands the field-of-view (FoV) of PMAS to a hexagonal area with a footprint of 74 x 65 arcsec$^{2}$. The observations have a final spatial resolution of full width at half maximum (FWHM) of $\sim3"$, corresponding to $\sim1$ kpc at the average redshift of the survey. The sampled wavelength range is 3745-7500 \AA, with a spectroscopic resolution of $\lambda$/$\bigtriangleup\lambda$ $\sim$850 for the low resolution setup, which is the one used in this work. The dataset was reduced using version 1.5 of the CALIFA pipeline. The galaxies used in this work are part of the 2nd CALIFA Data Release \citep[DR2, ][]{2015A&A...576A.135G}. The survey is already finished and all the galaxy datacubes are accesible from the 3rd Data Release webpage \footnote{http://califa.caha.es/DR3/} \citep[DR3, ][]{2016A&A...594A..36S}. Further details about the survey, sample, observational strategy and data reduction can be found in \cite{2012A&A...538A...8S,2013A&A...549A..87H,2014A&A...561A.130C,2014A&A...569A...1W} and references therein.

\subsection{Galaxy sample}
\label{Subsec: Galaxy sample}

Our galaxy sample is composed by 263 isolated, spiral galaxies observed by the CALIFA survey until September 2014. The sample was selected combining the isolated and merging classification and the morphological type designation from the CALIFA survey and the Hyperleda catalogue \citep{2014A&A...570A..13M}. It is fully characterised in Paper I, where it can be seen that their properties and parameter ranges fulfil the statistical properties of the whole CALIFA sample \citep[characterised itself in][hereafter W14]{2014A&A...569A...1W}, with the only exception of the exclusion of the earliest morphological types. Thus, the sample comprises galaxies of all spiral morphological types, with a 40.3\% of barred galaxies and a 17.1\% of galaxies where a ring can be observed. Also the redshift range of our galaxy sample covers the whole redshift range of the CALIFA mother sample, 0.005 < \textit{z} < 0.03. Distances to the galaxies in our work are obtained from the distance moduli given by the Hyperleda catalogue, which are corrected for Virgo-centric infall. They range from 21.9 to 129.2 Mpc, with a median value of 65 Mpc. 

Inclination values for the galaxy sample are derived using the \textit{b/a} axis ratios given by CALIFA and the expression given by \cite{1958MeLuS.136....1H} and first derived by \cite{1926ApJ....64..321H}, where the value of the axial ratio for an edge-on system parameter as a function of the galaxy morphological type is given by \cite{1972MmRAS..75...85H}. The inclination values distribution is skewed towards high inclination values, a result that reflects a selection effect already known to exist in the CALIFA mother sample and that specifically affects galaxies with low luminosity (see W14).

The effective radii of the galaxies, classically defined as the radius encompassing  half of the total light of the system, are used as normalisation factors in order to define, analyse and compare the inner and outer region samples as a function of their location in a given galaxy. We use the effective radii estimated by the CALIFA survey as described in \cite{2012A&A...546A...2S}, obtaining values between 1.63 and 20.87 kpc, which can be considered a wide range of sizes.  

Finally, the absolute magnitudes of our galaxy sample covers the range where the CALIFA sample is representative of the overall galaxy population (see W14).

\subsection{\hh region selection criteria}
\label{Subsec: region samples selection criteria}

The \hh region segregation and spectra extraction were performed using the {\sc HIIexplorer} procedure \citep{2012A&A...546A...2S,2012ApJ...756L..31R}. Subsequently, for each extracted spectrum, the stellar continuum was modelled using the {\sc FIT3D} package \citep{2006AN....327..167S,2011MNRAS.410..313S}. Individual emission line fluxes were then measured performing a multicomponent fitting using a single Gaussian function. A quality control process was applied to ensure a sample composed by physical \textit{bona fide} \hh regions and avoid selection uncertainties, establishing a valid range for the H$\alpha$/H$\beta$ line ratio and lower limits for the H$\alpha$ equivalent width and the H$\beta$ and \oiii~$\lambda$5007 lines fluxes. We finally obtained a sample of 9281 \hh regions. More details about these procedures can be found  in Paper I and references therein. From that total sample, we considered inner regions those that fulfilled the criterion given by \cite{2015MNRAS.451.3173A}:
\begin{equation}
\log {\rm R(kpc)} = - 0.204\cdot{\rm M_B} - 3.5. 
\end{equation}

Likewise, we considered as outer regions those that are located at a distance larger than two effective radii ($\textit{R}_{\rm eff}$) from the centre of their galaxy. The final work sample is composed by 725 inner regions and 671 outer regions.

\section{Functional parameters of the observed \hh regions samples}
\label{Sec: Functional parameters of the observed regions samples}

To first order, the emission line spectra of \hh regions are controlled by three factors: the shape of the ionising continuum, the degree of ionisation of the nebula and the abundance of the gas \citep{1991MNRAS.253..245D}. In simple \hh region models, they can be represented by functional parameters that can be  derived from observation: the effective temperature of the dominant ionising stars, T$_{\rm eff}$, the ionisation parameter, \textit{u}, and the oxygen abundance, 12 + log(O/H). In turn, these parameters are  related to physical properties of the ionising clusters: mass, metallicity and age,  and therefore are a powerful tool to study  differences in these physical characteristics. The determination of these functional parameter ranges in our observed samples provides inputs to the grid of the photoionisation models used for the interpretation of the data.

\subsection{Oxygen abundance}
\label{Subsec: Oxygen abundance}

The oxygen abundance, commonly expressed as 12 + log(O/H), can be used as a proxy for metallicity (total abundance of metals, \textit{Z}), assuming that $\log ({\rm Z/Z_{\odot}}) = \log {\rm (O/H)} - \log {\rm (O/H)_{\odot}}$. We consider as solar values Z$_{\odot}$ = 0.018 and $12 + \log {\rm (O/H)_{\odot}} = 8.69$ \citep{2009ARA&A..47..481A}.

The direct derivation of oxygen abundances from collisional excitation lines requires, at least, the detection and measurement of the auroral \oiii~$\lambda$4363 line for the electron temperature (T$_{e}$) to be determined. Unfortunately, this is not possible in CALIFA observations due to the limits set to the galaxy sample in terms of diameter and distance (see W14), except in the case of the nearest and most metal-poor galaxies. Therefore we have to rely on the observation of other stronger emission lines and the use of empirical calibrations. 

Currently, one of the most used indicators of oxygen abundance is the O3N2 index, firstly introduced by \cite{1979A&A....78..200A} and defined as
\begin{equation}
{\rm O3N2}=\log \left(\frac{{\rm [O\,III]}\,\lambda5007/H\beta}{{\rm [N\,II]}\,\lambda6583/H\alpha}\right).
\end{equation}

This index is based on the ratio of two of the strongest emission line observed and is weakly affected by differential extinction in comparison to other indicators that involve, for example, the \oii~$\lambda$3727 line. We have estimated the oxygen abundances of the inner and outer \hh region samples using the calibration obtained for this index by \cite{2013A&A...559A.114M}:
\begin{equation}
12 + \log {\rm (O/H)} = 8.533[\pm0.012] - 0.214[\pm 0.012]\cdot{\rm O3N2}.
\end{equation}

\begin{figure*}
\centering
\includegraphics[scale=0.5]{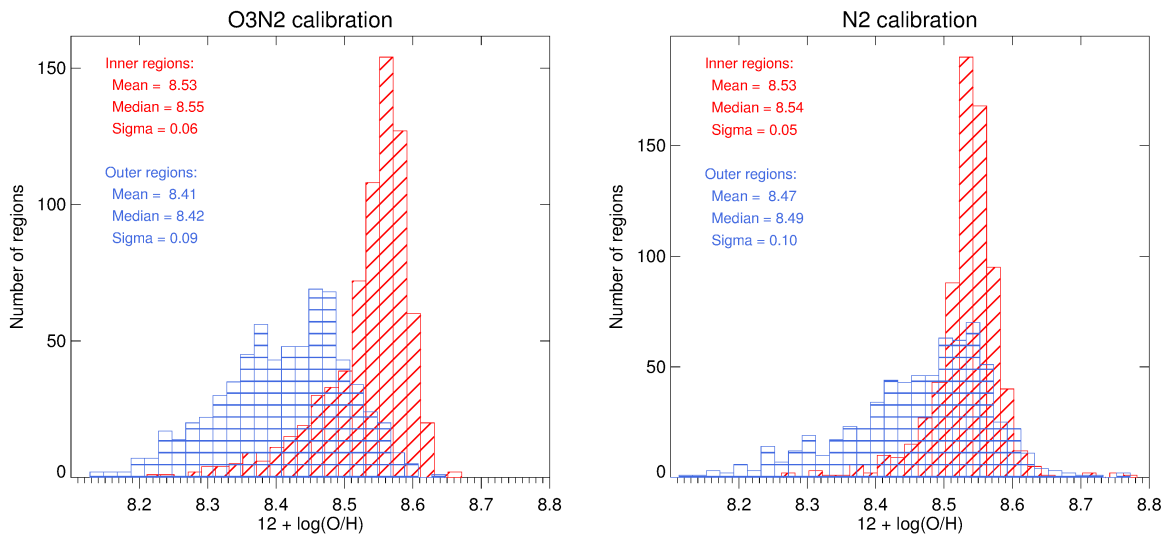}
\caption{Oxygen abundances derived using O3N2 index (left) and N2 index (right) for inner (red diagonally-hatched diagram) and outer (blue horizontally-hatched diagram) \hh region samples.}
\label{Fig: regions hist met calibrations}
\end{figure*}

This calibration is obtained by performing a linear fit in the range -1.1 < O3N2 < 1.7, corresponding to oxygen abundances between 8.17 and  8.77. The oxygen abundances obtained for both samples are represented in Fig. \ref{Fig: regions hist met calibrations}. As already mentioned in Paper I, we find somewhat higher abundances for the inner regions, ranging between 8.3 and 8.62 with a mean value of 8.53, while outer regions range between 8.2 and 8.6 with a mean value of 8.41. Although there is a certain overlapping between the metallicities of both samples, that of the inner regions is more picked. At any rate, none of them reaches solar or oversolar abundance values. This is in good agreement with the results obtained by previous works that focus on these kind of objects, which have found nebular oxygen abundances of about twice solar at most \citep{2002MNRAS.329..315C,2004ApJ...607L..21G,2005A&A...441..981B,2007MNRAS.382..251D}. In principle, this could be considered the result of some selection effects, mainly in the case of nebular abundances, since the higher the metallicity, the lower the gas electron temperature, and, therefore, the emission intensities of collisionally excited lines in the optical. However, observations of \hh regions in M51 in the infrared also find oxygen abundances close to twice solar \citep{2004ApJ...607L..21G}, and more recently studies of young stellar cluster abundances in M83 have found central abundances of \textit{Z} = $0.20\pm0.15$ \citep{2019ApJ...872..116H}. In fact, chemical evolution models predict a flattening of abundance gradients in spirals with metals reaching a saturation level \citep{2000MNRAS.313..338P,2005MNRAS.358..521M,2012A&A...540A..56P,2019MNRAS.482.3071M}.

Another indicator commonly used to determine the oxygen abundance is the N2 index, first proposed as calibrator by \cite{1994ApJ...429..572S} and defined as
\begin{equation}
{\rm N2}=\log \left(\frac{{\rm [N\,II]}\,\lambda6583}{{\rm H\alpha}}\right).
\end{equation}

It is based on a emission-line ratio very sensitive to the metallicity and independent of any reddening correction, due to the close wavelengths of the lines. \cite{2013A&A...559A.114M} also obtains a calibration for oxygen abundances based on this index:
\begin{equation}
12 + \log {\rm (O/H)} = 8.743[\pm0.027] + 0.462[\pm0.024]\cdot{\rm N2}.
\end{equation}

Although it saturates in the high-metallicity regime, N2 is found to correlate with abundance from low oxygen abundance values ($12 + \log {\rm (O/H)} = 7.8$; see Raimann et al. 2000) and up to $12 + \log {\rm (O/H)} = 9.2$ (Storchi-Bergmann et al. 1994). The calibration is computed in the range -1.6 < N2 < -0.2, which translates into an oxygen abundance range of $8.00 < 12 + \log {\rm (O/H)} < 8.65$. Some recent works based on CALIFA data, as \cite{2016A&A...585A..47M}, have used this calibration instead of the one based on the O3N2 index, considering that it provides a better match to the abundances obtained by T$_{e}$-based methods. We have calculated the oxygen abundances for both inner and outer region samples using this calibration, obtaining the results represented in the right panel of Fig. \ref{Fig: regions hist met calibrations}. It can be observed that the N2 calibration provides higher values of oxygen abundances, specially in the case of the outer regions. However, values are compatible within the errors considered in the calibrations, and the oxygen abundance range, which is the result we use as input in the photoionisation models parameter grid, is the same.

\subsection{Ionisation parameter}
\label{Subsec: Ionization parameter}

The ionisation parameter, \textit{u}, defined as the ratio of ionising photon density to hydrogen density, provides a measure of the degree of ionisation of the nebula. It can be expressed as:
\begin{equation}
u = \frac{\rm Q(H_0)}{4{\pi}{\rm c}{\rm n_H}{\rm {R_s}^{2}}},
\end{equation}

where $Q(H_0)$ is the number of ionising photons emitted by the ionising source, $R_s$ is the radius of the nebula, c is the speed of light and $n_H$ is the hydrogen density of the nebula. In this work the hydrogen density is considered constant throughout the nebula, and equal to the electron density for complete ionisation (case B of H recombination). 

In our scenario, we consider that the regions become observable in the optical range once they have dissipated their surrounding dust and the \hh regions have had time to reach a state of equilibrium with a constant nebula size $R_s$. Under these assumptions there is a direct relation between the ionisation parameter and the number of ionising photons. This implies a direct relation with the ionising mass of the cluster: for a given evolutionary stage, an \hh region with a larger ionising cluster mass will emit a larger number of ionising photons and thus will have a higher degree of ionisation. However, the number of ionising photons is also a decreasing function of the cluster age: as the cluster evolves, the most massive stars disappear and the number of ionising photons emitted to the region decreases. Therefore, the ionisation parameter is a function of the cluster mass and age, providing information about both physical parameters.

The ionisation parameter can be deduced from the ratio of two lines of the same element corresponding to two contigous ionisation states, such as the \oii~$\lambda$3727/\oiii~$\lambda$5007 or the \sii~$\lambda\lambda$6717,6731/\siii~$\lambda\lambda$9069,9532. In this case the \sii/\siii~ ratio cannot be used, as the \siii~ lines are out of our wavelength range. We do a first estimation of the ionisation parameter for inner and outer regions using the calibration given by \cite{2000MNRAS.318..462D}, based on the \oii/\oiii~ ratio:
\begin{equation}
\log {u} = - 0.80\cdot{\rm log (\oii/\oiii)} - 3.02.
\end{equation}

\begin{figure}
\centering
\includegraphics[scale=0.48]{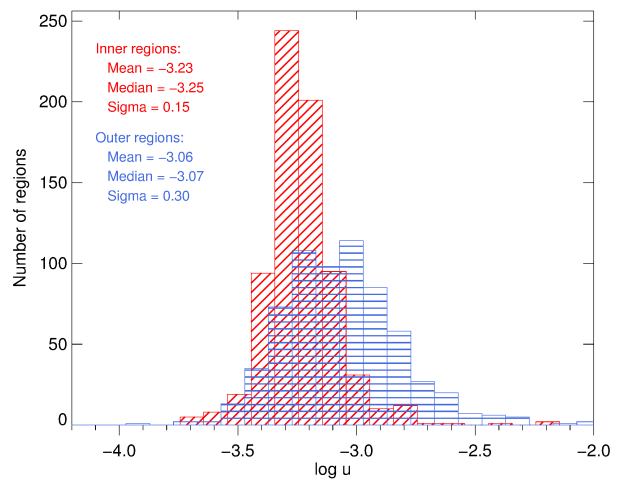}
\caption{Ionisation parameters derived for inner (red diagonally-hatched diagram) and outer (blue horizontally-hatched diagram) \hh region samples.}
\label{Fig: regions hist ionization parameter}
\end{figure}

We obtain ionisation parameter values ranging between -4.0 and -2.5 in logarithmic units, for both inner and outer regions. In order to gain consistency we have run a first set of models for this ionisation parameter range applying the same inputs that will be described in Sec. \ref{Sec: Photoionization models} and derived the relation between the input ionisation parameter and the \oii~$\lambda$3727/\oiii~$\lambda$5007 ratio, obtaining the expression:
\begin{equation}
\log {u} = - 0.75\cdot{\rm log (\oii/\oiii)} - 2.65,
\end{equation}
which is very similar to the one above. Applying this relation we have calculated the ionisation parameters for both inner and outer region samples, represented in Fig. \ref{Fig: regions hist ionization parameter}.

\subsection{Effective temperature}
\label{Subsec: Effective temperature}

The shape of the cluster ionising continuum is directly related to the effective temperature of the stars that dominate the radiation field responsible for the ionisation of the nebula \citep{2000MNRAS.318..462D},  which  depends on the cluster age and can be identified with the functional parameter $T_{\rm eff}$. This parameter can be related to the equivalent temperature of the clusters through the ratio of the number of helium ionising photons to the number of hydrogen ionising photons for a cluster of a given age and metallicity.

The stellar effective temperature is the most difficult parameter to derive observationally, since there is not a suitable emission line ratio that, by itself or used in combination with other, provides a good indicator. In the case of extragalactic regions, methods have been developed to estimate the $T_{\rm eff}$ through the analysis of the spectral emission lines \citep{1988MNRAS.231..257V,2003A&A...404..969D,2011ApJ...738...34P}. However, the influence of the gas abundance and the ionisation parameter on the emission line spectra, in the wavelength range used in this work,  prevents an unequivocal determination of the $T_{\rm eff}$. Several studies \citep{2000ApJ...537..589K,2004ApJ...601..858M} have reported the strong degeneracy existing between the effects of these parameters, with the consequent difficulty in the use of emission lines as $T_{\rm eff}$ indicators. 

In order to break this degeneracy, \cite{2017MNRAS.466..726D} have produced a calibration between the \oii~$\lambda\lambda$3727,3729/\oiii~$\lambda$5007 emission-line ratio and the effective temperature of \hh regions for different values of the ionisation parameter (see Fig. 2 and Table 2 in the mentioned paper). In our case this method is degenerated, since it would imply using the same emission-line ratio as proxy for both \textit{u} and $T_{\rm eff}$. But as a first qualitative approach, considering the calibrations that they provide for solar and half-solar metallicities and ionisation parameters between -2.5 and -3.5 in logarithmic units, we obtain that most of our regions have $T_{\rm eff}$ higher than 40000 K. Taking into account the uncertainty in the ionisation parameter estimations, we consider that it is only possible to derive a lower limit of 37000 K for the effective temperature of our regions, as more precise values cannot be taken with confidence.

\section{Photoionisation models}
\label{Sec: Photoionization models}

\subsection{Parameter grid}
\label{Subsec: Parameters grid}


We have used the photoionisation code Cloudy \citep[version 17.01, last described by][]{2017RMxAA..53..385F} to simulate the emission-line spectra of \hh regions with functional parameters similar to those shown by our observational samples. All our models have been computed assuming an ionisation-bounded geometry, with the emitting gas located in a thin spherical shell at a distance from the ionising source large enough, when compared to the shell thickness, to validate the approximation of a plane-parallel geometry. 


We have considered as ionising sources coeval star clusters whose spectral energy distributions (SEDs) are given by PopStar evolutionary synthesis models \citep{2009MNRAS.398..451M,2010MNRAS.403.2012M}, considering a Salpeter initial mass function \citep[IMF,][]{1955ApJ...121..161S} with lower and upper mass limits of 0.15 and 100 M$_{\odot}$ respectively. Ionising clusters input metallicities are \textit{Z} = 0.006, 0.009, 0.012, 0.015 and 0.018 (selected as explained further in this section), constructed by interpolation in the PopStar tables as done by the Cloudy routine. The value \textit{Z} = 0.018 corresponds to solar metallicity, according to \cite{2009ARA&A..47..481A}.

We have followed the evolution of the simulated clusters from ages 1 to 5.2 Myr. The starting age reflects the fact that the observed \hh regions are not in the very first phases of their evolution, since regions younger than that age tend to be heavily obscured, with their ionising clusters still embedded in their initial molecular cloud, and therefore are not visible at optical wavelengths \citep{2006ApJS..167..177D}. On the other hand, a final computing age of 5.2 Myr has been chosen since star clusters older than this do not produce the necessary hydrogen ionising photons as to produce a detectable emission-line spectrum \citep{2010MNRAS.403.2012M}.


The hydrogen density is considered constant throughout the nebula, and thus equal to the electron density for complete ionisation. The \sii~$\lambda$6717/\sii~$\lambda$6731 for both inner and outer \hh region samples show similar average values (see Paper I), corresponding to electron densities smaller than 10 cm$^{-3}$ \citep{2006agna.book.....O}. We therefore compute models with an hydrogen density fixed value of $n_{\rm H}$ = 10 cm$^{-3}$, that is, in the low density regime.


We need to introduce the chemical composition of the gas as an input to the photoionisation code. Here we have assumed that gas and ionising stars have the same metallicity, since the massive stars do not live long enough as to travel very far from the gas from which they formed. We have also assumed that the medium is chemically homogeneous. As mentioned above, solar abundance values have been taken from \cite{2009ARA&A..47..481A}, which implies that the solar metallicity is \textit{Z}$_{\odot}$ = 0.018 and the solar oxygen abundance is $12 + \log {\rm (O/H)_{\odot}} = 8.69$. We have also applied depletion factors for some refractory elements (Na, Mg, Al, Si, Ca, Fe and Ni), as a certain proportion of these elements can be part of dust grains that are mixed with the ionised gas \citep{1995ApJ...449L..77G}. Considering the range of oxygen abundances obtained for the observed \hh regions on Sec. \ref{Subsec: Oxygen abundance}, and assuming oxygen abundances as a direct tracer of metallicity, we have run models with \textit{Z} = 0.006, 0.009, 0.012, 0.015 and 0.018.

The elemental abundances relative to oxygen have been assumed solar except in the case of nitrogen, for which we have taken into account the primary and secondary production mechanisms. The latter one becomes important in the case of gas with relatively high abundances, that is, when the oxygen abundance is 12 + log(O/H) > 7.5-8.0 \citep[see, e.g.][]{2006MNRAS.372.1069M} which, as shown in Sec. \ref{Subsec: Oxygen abundance}, is the case for our observed \hh region samples. In order to check this behaviour, we have studied the relation between the N/O ratio and the oxygen abundance in our data. In fact, there is a good correlation between the \nii~$\lambda$6583/\oii~$\lambda$3727 emission-line ratio and the N$^{\textit{+}}$/O$^{\textit{+}}$ ionic abundance ratio, which traces the nitrogen to oxygen abundance ratio \citep{2005MNRAS.361.1063P,2009MNRAS.398..949P}. According to \cite{2007MNRAS.382..251D}, this relation can be expressed by:
\begin{equation}
\log {\rm (N/O)} = 0.65\cdot\log ([\ion{N}{ii}]/[\ion{O}{ii}]) - 0.79.
\end{equation}
 
\begin{figure*}
\centering
\includegraphics[scale=0.5]{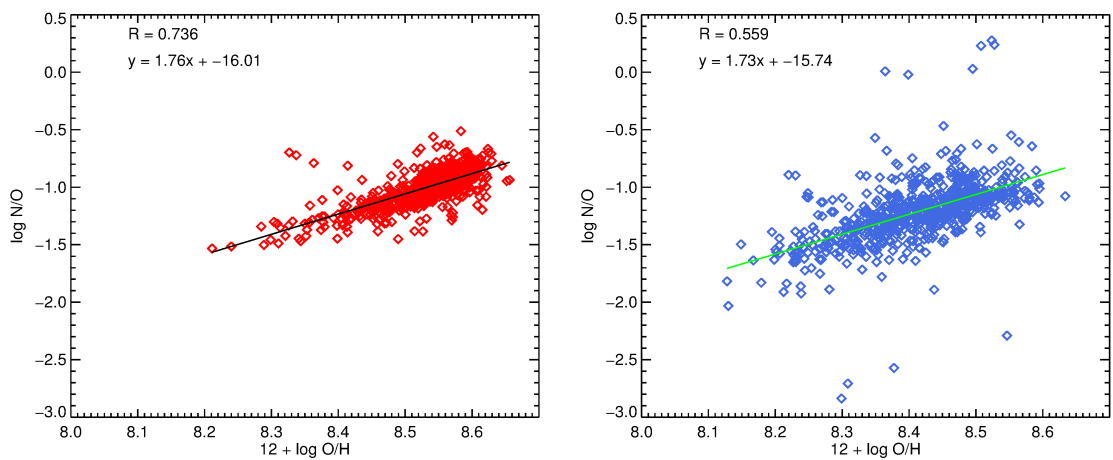}
\caption{Relation between the N/O and the O/H abundance ratios for inner (left, red) and outer (right, blue) observed \hh region samples.}
\label{Fig: N/O vs 12+logO/H}
\end{figure*}

Using this relation we have derived the N/O abundances for the studied regions. Fig. \ref{Fig: N/O vs 12+logO/H} shows the N/O ratio as a function of the oxygen abundaces already derived in Sec. \ref{Subsec: Oxygen abundance} for both inner and outer region samples, and the linear fitting obtained in each case. The influence of the secondary contribution of nitrogen is clearly observed. We use the regression fitting obtained for the inner region sample, which has a slightly higher regression coefficient, to include the dependence of the N/O relative abundances on the O/H abundance that traces metallicity in the Cloudy models. 

The considered solar abundances and the final depleted ones, as well as the N/H abundances obtained applying the N/O abundances derived for each metallicity, are summarised in Table \ref{Table: abundances} for all the metallicity input values computed in the models.

The ionisation parameter is also introduced as an input to the models. We have used the range of values from -3.75 to -2.5 in logarithmic units, estimated from the observational samples as mentioned in Sec. \ref{Subsec: Ionization parameter}, with steps of 0.25 logarithmic units. Therefore we have computed models with log \textit{u} = -3.75, -3.5, -3.25, -3.0, -2.75 and -2.5.  

We have run models at 18 steps of time, log [t (yr)] = 6.00, 6.10, 6.18, 6.24, 6.30, 6.35, 6.40, 6.44, 6.48, 6.51, 6.54, 6.57, 6.60, 6.63, 6.65, 6.68, 6.70 and 6.72. A total of 540 photoionisation models have been computed, which results and comparison to the observational samples are presented in the following sections.

\begin{table*}
\caption{Summarised solar and depleted abundances for each metallicity used in the Cloudy photoionisation models.}  
\label{Table: abundances}    
\centering                                    
\begin{tabular}{l c c c c c c }         
\hline\hline                       
\noalign{\smallskip}              
Element & Solar & 0.006 & 0.009 & 0.012 & 0.015 & 0.018 (solar) \\
 &  & depleted & depleted & depleted & depleted & depleted \\

\hline

\noalign{\smallskip}

He & -1.07 & -1.07 & -1.07 & -1.07 & -1.07 & -1.07 \\
C   & -3.57 & -4.04 & -3.87 & -3.74 & -3.64 & -3.57 \\
N   & -4.03 & -5.34 & -4.85 & -4.51 & -4.24 & -4.03 \\
O   & -3.31 & -3.78 & -3.61 & -3.48 & -3.38 & -3.31 \\
Ne & -4.07 & -4.54 & -4.37 & -4.24 & -4.14 & -4.07 \\
Na & -5.76 & -7.23 & -7.06 & -6.93 & -6.83 & -6.76 \\
Mg & -4.40 & -5.87 & -5.70 & -5.57 & -5.47 & -5.40 \\
Al  & -5.55 & -7.02 & -6.85 & -6.72 & -6.62 & -6.55 \\
Si  & -4.49 & -5.26 & -5.09 & -4.96 & -4.86 & -4.79 \\
S   & -4.88 & -5.35 & -5.18 & -5.05 & -4.95 & -4.88 \\
Ar & -5.60 & -6.07 & -5.90 & -5.77 & -5.67 & -5.60 \\
Ca & -5.66 & -7.13 & -6.96 & -6.83 & -6.73 & -6.66 \\
Fe & -4.50 & -5.97 & -5.80 & -5.67 & -5.57 & -5.50 \\
Ni & -5.78 & -7.25 & -7.08 & -6.95 & -6.85 & -6.78 \\

\hline
\end{tabular}  
\end{table*}

\subsection{Diagnostic diagrams}
\label{Subsec: Diagnostic diagrams}

Emission-line diagnostic diagrams are used to explore the relation between different strong emission-line ratios, which can provide information about the physical conditions of the ionised gas and the nature and physical properties of the ionising sources. In this case we study the predictions of the computed photoionisation models in relation to the location of the observational samples in diagrams based on emission-line ratios related to properties as oxygen abundance, ionisation parameter or ionising cluster age. Divergences in trends or parameter ranges identified for inner and outer regions may provide important information about differences in evolutionary stage, metal content or total mass of the regions as a function of their location and environment.

\subsubsection{Classic BPT diagram: \oiii~$\lambda$5007/H$\beta$ vs \nii~$\lambda$6583/H$\alpha$}
\label{Subsubsec: Diagnostic diagram: nitrogen}

One of the classic diagnostic diagrams \citep[introduced by][hereafter BPT]{1981PASP...93....5B} is the one that represents \oiii~$\lambda$5007/H$\beta$ vs \nii~$\lambda$6583/H$\alpha$. These two emission-line ratios have the advantage of using lines at relatively close wavelengths, thus reducing or avoiding the effects of differential extinction. 

\begin{figure}
\centering
\includegraphics[scale=0.48]{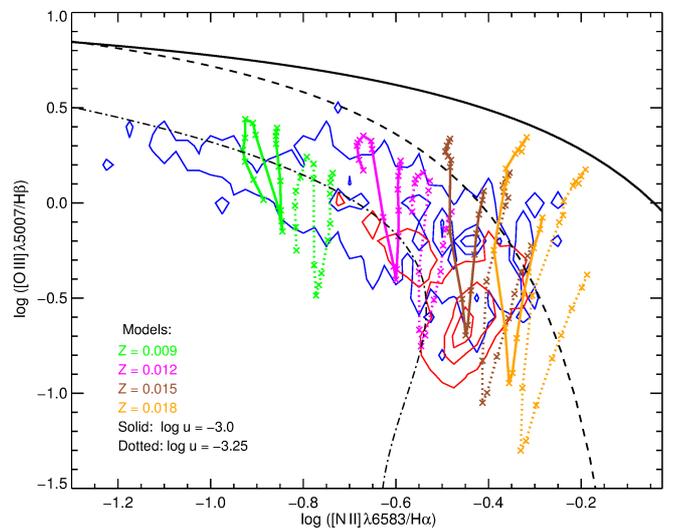}
\caption{Diagnostic diagram representing \oiii~$\lambda$5007/H$\beta$ vs \nii~$\lambda$6583/H$\alpha$. Observed inner (red) and outer (blue) \hh regions distributions are represented with density contours. Photoionisation models predictions are overplotted: colours indicate different input metallicities, and solid and dotted lines indicate different input ionisation parameters, as showed in the plot legend. The solid and dotted black lines represent the Kewley et al. (2001) and Kauffmann et al. (2003) demarcation curves, respectively. The dash-dotted black line represents the fit made by Rousseau-Nepton et al. (2018) for NGC\,628, as explained in the text.}
\label{Fig: BPT [OIII]/Hb vs [NII]/Ha}
\end{figure}

Fig. \ref{Fig: BPT [OIII]/Hb vs [NII]/Ha} shows the location of inner and outer \hh regions in this diagram, using density contours in order to better observe the distribution trends shown by the data. Cloudy photoionisation model predictions are represented for different values of the model input metallicity, including only \textit{Z} = 0.009, 0.012, 0.015 and 0.018 models, since \textit{Z} = 0.006 models were clearly not matching the location of the observational data and could easily be ruled out. For each metallicity we represent models for two different ionisation parameters: log \textit{u} = -3.0, -3.25, which are the ones closer to the mean and median values of log \textit{u} derived for each observational region sample (see Fig. \ref{Fig: regions hist ionization parameter}). The line of every model corresponds to the evolving age of the ionising cluster, from 1 to 5.2 Myr.

As already discussed in Paper I, in this diagram, inner \hh regions are located to the bottom right of the classical star forming branch, implying higher oxygen abundances and lower excitations than those shown by  outer \hh regions. It is important to note that inner regions are mostly located at the vertical zone of the star-forming branch, implying that they may be suffering the effect of the \nii~saturation at close to solar metallicities. In the case of photoionisation models, we observe the expected trend indicated by the star forming branch, with models being displaced to the right of the diagram for increasing input metallicities. When a fixed input metallicity is considered, models with a lower value of the input ionisation parameter are progressively located to the bottom-right zone of the diagram. 

The diagram also includes the polynomial fit obtained by \cite{2018MNRAS.477.4152R} for 4285 \hh regions from the disk of NGC\,628, observed with the CFHT imaging spectrograph SITELLE. The location of this fit depends on the metallicity of the regions used in its derivation and this is a specific result for NGC\,628, so we do not expect it to reproduce our sample distributions, since they are composed by regions belonging to a wide variety of galaxies. But interestingly enough it shows a turnover at log (\oiii~$\lambda$5007/H$\beta$)$\sim$-0.5 and log (\nii~$\lambda$6583/H$\alpha$)$\sim$-0.5 that we do observe in our inner \hh region sample. This turnover may be expected for metallicities above 12+log[O/H]$\sim$9 while considering ionisation parameters decreasing with increasing metallicity, but it could also be associated to a variation in the relative abundances, as the N/O abundance \citep{2018MNRAS.477.4152R}. 

The qualitative comparison of observational data and photoionisation models in this diagram provides information about the metallicity of the observed \hh regions and, to a lesser extent, about their ionisation parameters, and confirms some expected trends for inner and outer regions. At the lowest represented metallicity (\textit{Z} = 0.009), models with the highest values of ionisation parameter seem to reconstruct the trend shown by the top left tail of the observed outer regions, while only a few inner regions seem to be possibly associated to these values of metallicity and ionisation parameter. On the contrary, models with \textit{Z} = 0.012 and \textit{Z} = 0.015 seem to reconstruct the trend shown by both region samples, with lower ionisation parameters in the case of the inner \hh regions. Models with the highest computed metallicity, \textit{Z} = 0.018, can be associated with inner and outer regions with the highest \nii~$\lambda$6583/H$\alpha$ values of their corresponding samples, although high values of ionisation parameters would be required in this case. In general, models confirm higher metallicities and lower ionisation parameters for the inner regions. Nevertheless, the \nii~saturation at close to solar metallicities may be affecting the inner regions, avoiding a quantitative determination of their metallicities. Therefore the use of the \nii~$\lambda$6583/H$\alpha$ ratio as oxygen abundance tracer may produce an underestimation of the inner regions oxygen abundances, and the range of reliability of calibrations based on this emission-line ratio should be carefully reconsidered when inner regions are involved.

The strong degeneracy of the models with age, which affects specifically the \oiii~$\lambda$5007/H$\beta$ emission-line ratio, prevents any conclusion about the ages of the ionising clusters. This degeneracy is caused by the appearance of the WR at ages of around 2.5-3 Myr, producing a temporal hardening of the ionising radiation. Assuming an instantaneous burst, WR features are present in the cluster spectra from ages around 3 Myr to almost 7 Myr, with shorter lifetimes for lower metallicities \citep{1998ApJ...497..618S}, as it is observed in the represented models.  

\subsubsection{Evolution of \sii~$\lambda$6717,6731/H$\alpha$}
\label{Subsubsec: Diagnostic diagram: sulphur}

\begin{figure}
\centering
\includegraphics[scale=0.48]{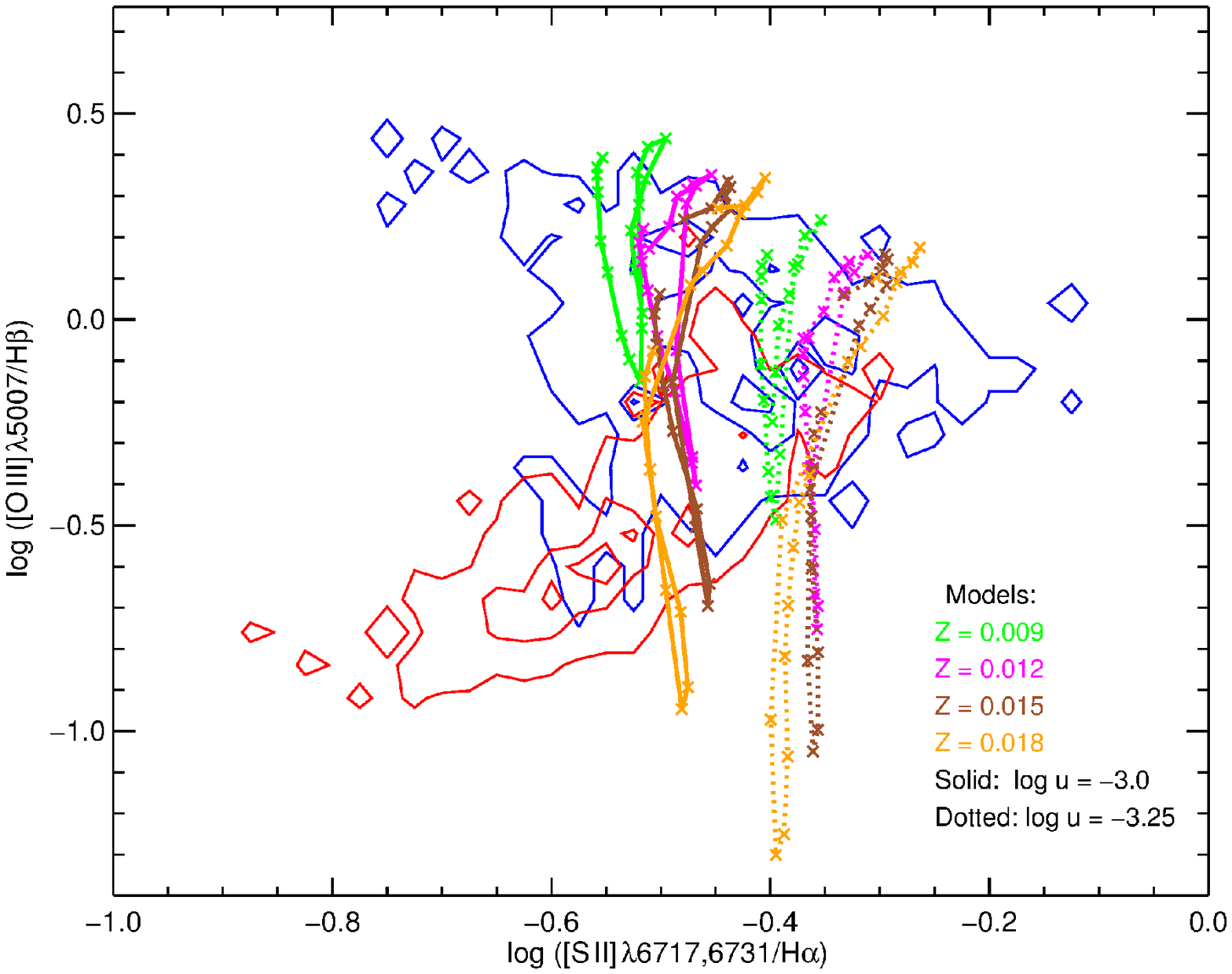}
\caption{Diagnostic diagram representing \oiii~$\lambda$5007/H$\beta$ vs \sii~$\lambda$6717,6731/H$\alpha$. Observed inner (red) and outer (blue) \hh regions distributions are represented with density contours. Photoionisation models predictions are overplotted: colours indicate different input metallicities, and solid and dotted lines indicate different input ionisation parameters, as showed in the plot legend.
}
\label{Fig: BPT [OIII]/Hb vs [SII]/Ha}
\end{figure}

Another classical and widely used diagnostic diagram is the one representing \oiii~$\lambda$5007/H$\beta$ versus \sii~$\lambda$6717,6731/H$\alpha$, firstly introduced by \cite{1987ApJS...63..295V}. The \sii/H$\alpha$ ratio also avoids a strong influence of reddening due to the close wavelengths of the emission lines involved, and has the advantage of being independent of relative abundances sinces both oxygen and sulphur have the same nucleosynthetic history. Fig. \ref{Fig: BPT [OIII]/Hb vs [SII]/Ha} shows the distribution of the observed \hh region samples in this diagram using density contours and, although the samples locations partially overlap, different trends can be observed. The characteristic turnover shown by the star-forming branch in this diagram at log (\oiii~$\lambda$5007/H$\beta$) $\sim$ 0.0 and log (\sii~$\lambda$6717,6731/H$\alpha$) $\sim$ -0.5 \citep[see e.g.][]{2006MNRAS.372..961K}  is clearly seen for both observed samples. Outer regions are mostly located in the turnover region and in the upper branch, with \sii~$\lambda$6717,6731/H$\alpha$ increasing as \oiii~$\lambda$5007/H$\beta$ decreases, while inner regions are mostly located in the lower branch, that shows a decrease of \sii~$\lambda$6717,6731/H$\alpha$ with the decreasing \oiii~$\lambda$5007/H$\beta$. This last behaviour is also found for \hh regions in high-metallicity galaxies, as the ones observed in NGC\,5194 (also called M51a) by \cite{2015ApJ...808...42C}. 

The predictions by Cloudy photoionisation models are also represented in Fig. \ref{Fig: BPT [OIII]/Hb vs [SII]/Ha}, following a similar scheme to the one indicated in Fig. \ref{Fig: BPT [OIII]/Hb vs [NII]/Ha}. Regarding metallicity, models reproduce the turnover for increasing \textit{Z}, but it is not possible to clearly discriminate the metallicity of the observed regions, due to the proximity of the model predictions for different \textit{Z}. On the other hand, we do observe an evolution with the ionisation parameter, as models displace to the bottom-right part of the diagram with decreasing values of log \textit{u}. This reproduces the upper branch of the diagram, populated by the outer regions, which are well fitted by models with log \textit{u} = -3.0 and log \textit{u} = -3.25, as expected. On the contrary, there is a considerable amount of inner regions located at the end of the lower branch that do not seem to be associated to any model predictions, not even in the case of the models with solar metallicity. 

\begin{figure*}
\centering
\includegraphics[scale=0.5]{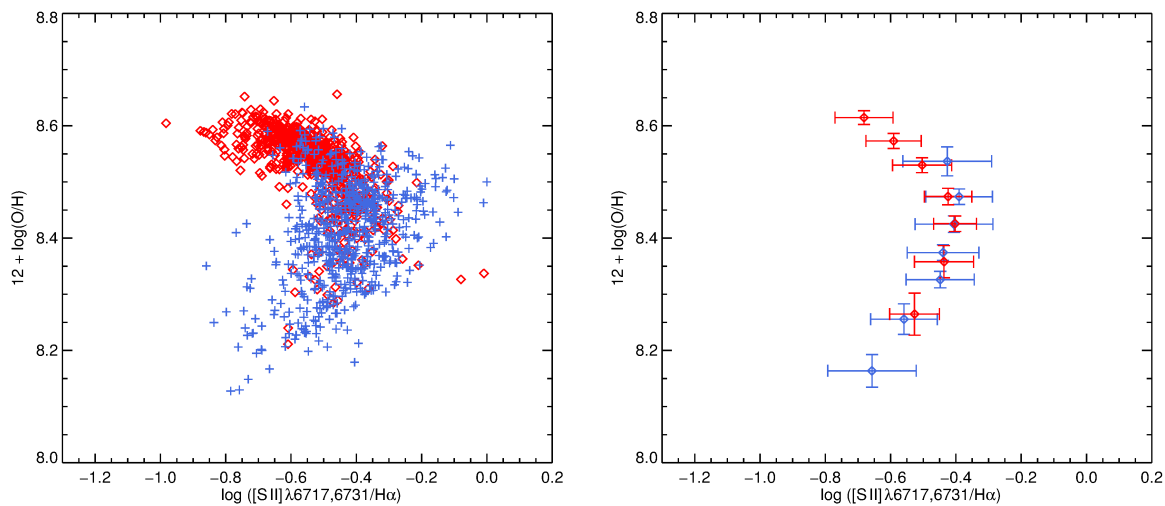}
\caption{Diagnostic diagram representing the oxygen abundance, 12 + log(O/H), vs \sii~$\lambda$6717,6731/H$\alpha$. Left: Inner (red diamonds) and outer (blue crosses) observed \hh regions samples. Right: Average values of the represented line ratios considering bins of oxygen abundance for inner (red) and outer (blue) observed \hh regions.
}
\label{Fig: Oxygen abundance vs [SII]/Ha}
\end{figure*}

This behaviour of the \sii~$\lambda$6717,6731/H$\alpha$ observational data leads us to consider the implications this may have in the relation between this ratio and the metallicity. Left panel of Fig. \ref{Fig: Oxygen abundance vs [SII]/Ha} represents the oxygen abundance values obtained using the O3N2 calibration (see Sec.  \ref{Subsec: Oxygen abundance}) versus \sii/H$\alpha$ for the inner and outer \hh region samples. We have also considered bins of 0.1 in 12 + log(O/H) and calculated average values for the two magnitudes involved in this plot for both inner and outer region samples. These average values per bin are represented in the right panel of Fig. \ref{Fig: Oxygen abundance vs [SII]/Ha}. The double-valued relation of this ratio with the oxygen abundance is clearly seen for both samples, with the turnover region at 12 + log(O/H) $\sim$ 8.45. This result strongly affects the range of reliability of ionisation parameter calibrations based on this ratio and metallicity \citep[e.g.][]{2000MNRAS.318..462D}, which should only be considered valid for metallicities up to the turnover point, and not to be applied when inner regions are considered.

\subsubsection{\nii~$\lambda$6583/\oii~$\lambda$3727 versus O3N2}
\label{Subsubsec: Diagnostic diagram: NII/OII vs O3N2}

The diagnostic diagram representing \nii~$\lambda$6583/\oii~$\lambda$3727 vs the O3N2 index also provides interesting information about the data, as already seen in Paper I, Fig. 14. The \nii/\oii~ line ratio was introduced as an abundance indicator by \cite{2000ApJ...542..224D}, due to two main reasons. The first one is that the secondary production mechanism of nitrogen makes this ratio to systematically increase with abundance over a wide range of metallicity values, including the ones shown by our regions, as explained in Sec. \ref{Subsec: Parameters grid}. The second one is related to the different wavelengths of the emission lines involved and thus to the distance between the location of the lines in the spectrum: when the metallicity increases, the mean electron temperature of the \hh regions decreases, and at high metallicities this temperature becomes too low to excite transitions with high energy as the \oii~$\lambda$3727, while the \nii~$\lambda$6583 continues to be excited. Thus, the \nii/\oii~ ratio increases with metallicity. 

\begin{figure}
\centering
\includegraphics[scale=0.48]{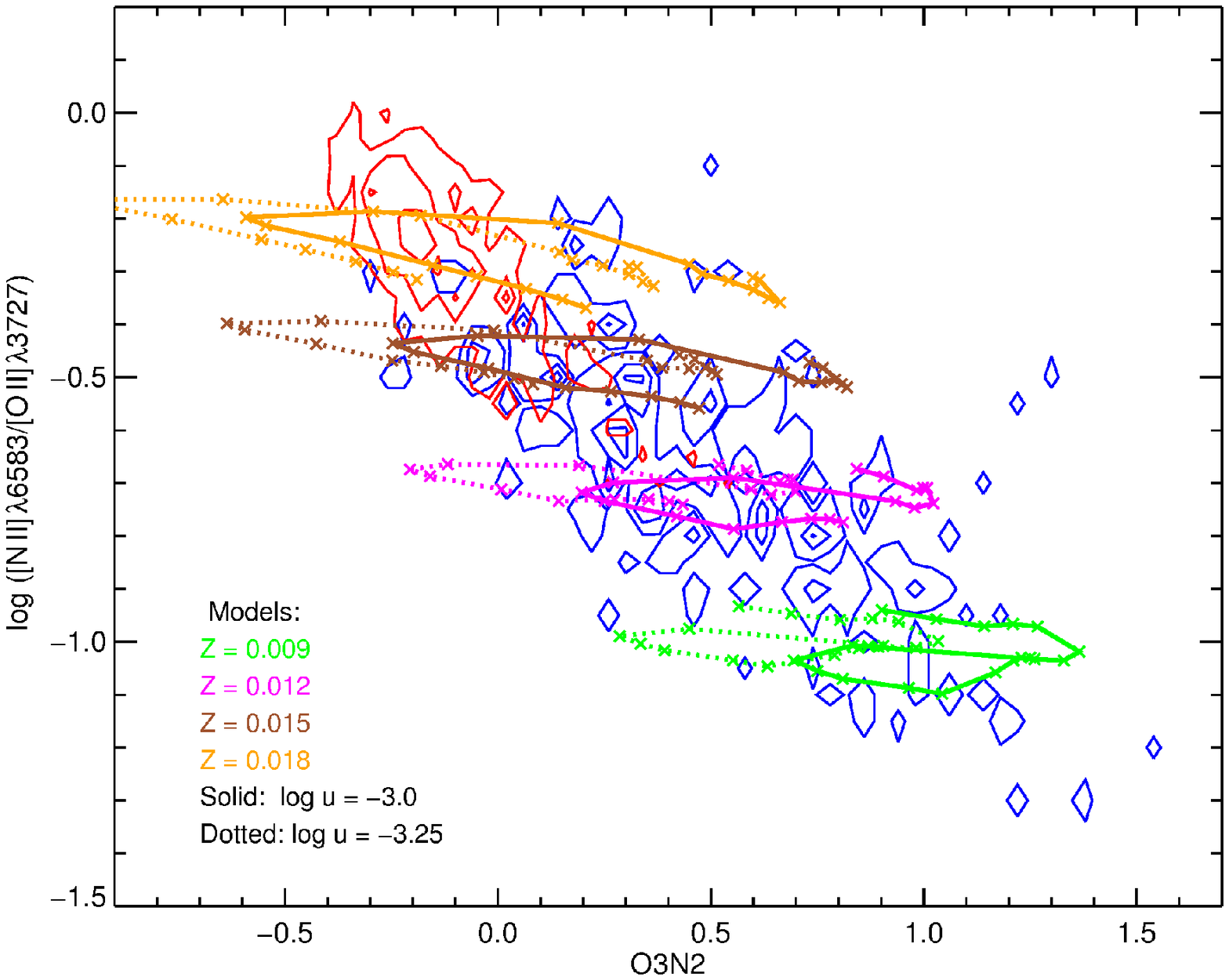}
\caption{Diagnostic diagram representing \nii~$\lambda$6583/\oii~$\lambda$3727 vs the O3N2 index. Observed inner (red) and outer (blue) \hh regions distributions are represented with density contours. Photoionisation models predictions are overplotted: colours indicate different input metallicities, and solid and dotted lines indicate different input ionisation parameters, as showed in the plot legend.
}
\label{Fig: BPT [NII]/[OII] vs O3N2}
\end{figure}

The distributions of inner and outer \hh region samples in this diagram are represented with colour contours in Fig. \ref{Fig: BPT [NII]/[OII] vs O3N2}. The predictions from photoionisation models are also plotted, following the same scheme employed in Fig. \ref{Fig: BPT [OIII]/Hb vs [NII]/Ha} and Fig. \ref{Fig: BPT [OIII]/Hb vs [SII]/Ha}. It can be observed that the models reproduce the trend of the data, with higher metallicities for increasing \nii/\oii~ and decreasing O3N2 index, as expected. Outer regions are fitted by models with metallicities between 0.009 and 0.015. It is important to note the strong degeneracy existent in the model predictions between age and ionisation parameter, as for a given metallicity the variation of both parameters produces a horizontal displacement of the predicted values. In the case of the ionisation parameter, lower values of log \textit{u} displace models to lower values of O3N2. Regarding age, the effect varies with increasing time, depending on the hardening of the spectra produced by the WR appearance, as mentioned in Sec. \ref{Subsubsec: Diagnostic diagram: nitrogen}. This degeneracy has two important consequences. In the first place, it clearly prevents the discrimination of \hh regions ages and ionisation parameters by comparison with models in this diagram. Secondly, it indicates that even though we are relating line ratios strongly dependent of metallicity and it is possibly to observe a relation between them, the scatter observed in the data is intrinsic, due to their physical conditions (age and ionisation structure), and not observational. This implies that any empirical calibration made for the oxygen abundance based on the O3N2 index can be strongly influenced by the physical properties of the \hh region sample employed, and no univocal relation can be established without characterising these properties. The effect of the ionisation parameter in the scatter of the O3N2 relation has been previously discussed by \cite{2005MNRAS.361.1063P}, \cite{2010IAUS..262...93S} and \cite{2010A&A...517A..85L}.

Inner regions are partly fitted by models with the highest computed metallicity, \textit{Z} = 0.018 (solar), or are located above the predictions of the models, in contrast to the \textit{Z} = 0.015 metallicity of the best-fitting models in the \oiii~$\lambda$5007/H$\beta$ vs \nii~$\lambda$6583/H$\alpha$ diagram. This might be partly caused by an underestimation of the oxygen abundance due to the \nii~saturation in that diagram, as discussed in Sec. \ref{Subsubsec: Diagnostic diagram: nitrogen}, but it can also be produced by an enhancement of the N/O abundance ratio values above the ones expected considering a linear relation between N/O and O/H, thus indicating an increase in the relation slope for the physical conditions of the inner regions. Previous studies have also found evidence of a non-linear relation between N/O and O/H, with a steeper relation between abundance ratios for higher oxygen abundances \citep{2000ApJ...541..660H}, and observations show that high-metallicity \hh regions have N/O ratios above the trend of N/O versus O/H that marks the secondary behaviour of nitrogen \citep[e.g.][]{2004ApJ...615..228B,2007MNRAS.382..251D,2015ApJ...808...42C}. This effect should be further characterised in order to be taken into account when developing models to study physical characteristics of circumnuclear star-forming regions (CNSFR).

\subsubsection{\oii~$\lambda$3727/\oiii~$\lambda$5007 versus O3N2}
\label{Subsubsec: Diagnostic diagram: OII/OIII vs O3N2}

\begin{figure}
\centering
\includegraphics[scale=0.48]{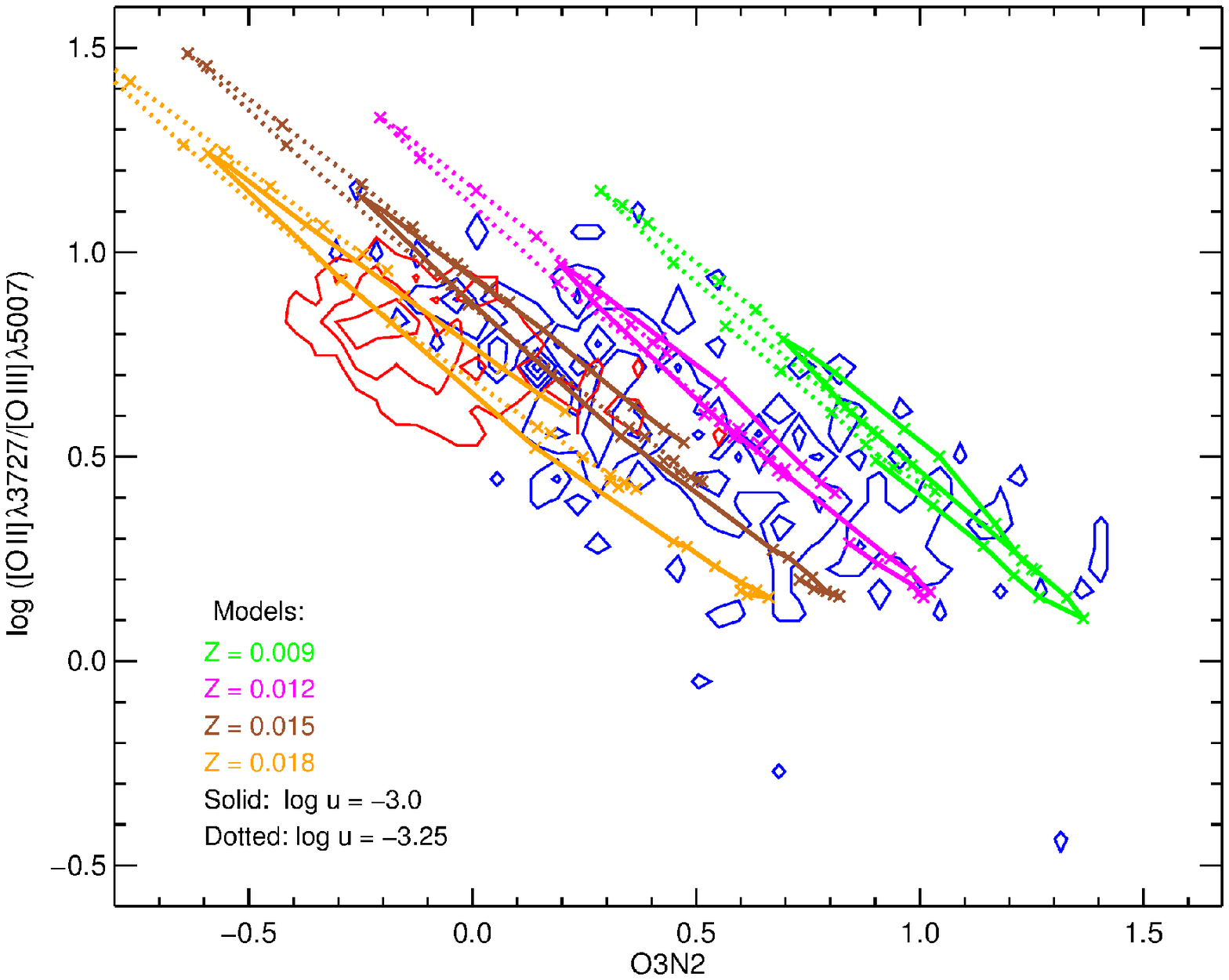}
\caption{Diagnostic diagram representing \oii~$\lambda$3727/\oiii~$\lambda$5007 vs the O3N2 index. Observed inner (red) and outer (blue) \hh regions distributions are represented with density contours. Photoionisation models predictions are overplotted: colours indicate different input metallicities, and solid and dotted lines indicate different input ionisation parameters, as showed in the plot legend.}
\label{Fig: BPT [OII]/[OIII] vs O3N2}
\end{figure}

As explained in Sec. \ref{Subsec: Ionization parameter}, the \oii~$\lambda$3727/\oiii~$\lambda$5007 emission-line ratio is sensitive to the ionisation parameter of the nebula. Therefore, representing this ratio versus the O3N2 index allows us to relate a proxy for the ionisation parameter and a proxy of metallicity, and provides an insight on the relation between these two functional parameters. Inner and outer  \hh regions are represented with contour plots in this diagram in Fig. \ref{Fig: BPT [OII]/[OIII] vs O3N2}, along with model predictions in a similar way to previous diagnostic diagram figures. 

Models reproduce the trend shown by the data also in this case, although with certain peculiarities. Models move to lower values of the O3N2 index for increasing metallicities, with a soft displacement to higher values of the \oii~$\lambda$3727/\oiii~$\lambda$5007 ratio. For a given metallicity, a decrease in the model ionisation parameter produces a displacement of the predictions to lower values of the O3N2 index and higher values of the \oii~$\lambda$3727/\oiii~$\lambda$5007 ratio, following the line defined by the evolution of the model predictions with age. We can reach two conclusions: first, we are able to partially discriminate between different metallicities of the data, but again the influence of the ionisation parameter and the age of the regions can affect this result, and should be taken into account in any calibration between the O3N2 index and the oxygen abundance, as discussed in Sec. \ref{Subsubsec: Diagnostic diagram: NII/OII vs O3N2}. Secondly, we are observing again a strong degeneracy between the effects produced by different ages and ionisation parameters, which prevents us from any precise differenciation of these two magnitudes using this diagram.

Keeping this in mind, outer \hh regions are fitted by models with metallicities between \textit{Z} = 0.009 and \textit{Z} = 0.015, as expected, and to a great extent they are fitted by models with log \textit{u} = -3.0. On the other hand, inner regions are fitted by models with higher metallicities: \textit{Z} = 0.015 or \textit{Z} = 0.018 (solar). However, a considerable amount of the inner regions are not covered by models and show a trend to lower \oii~$\lambda$3727/\oiii~$\lambda$5007 ratios than expected considering model predictions and the linear trend indicated by the outer regions. This may be caused by two possible reasons: either the \oii~$\lambda$3727/\oiii~$\lambda$5007 does not trace the ionisation parameter for the inner regions physical conditions, or the relation between ionisation parameter and oxygen abundance is not linear, and it changes for the inner regions physical conditions. As in the case of the N/O enhancement in Sec. \ref{Subsubsec: Diagnostic diagram: NII/OII vs O3N2}, this situation should be further characterised in order to be considered when inner regions and/or CNSFRs are studied.

\section{Evolutionary stages and masses}
\label{Sec: Evolutionary stages}

\subsection{Equivalent widths and composite populations}
\label{Subsec: Equivalent widths and composite populations}

\begin{figure}
\centering
\includegraphics[scale=0.48]{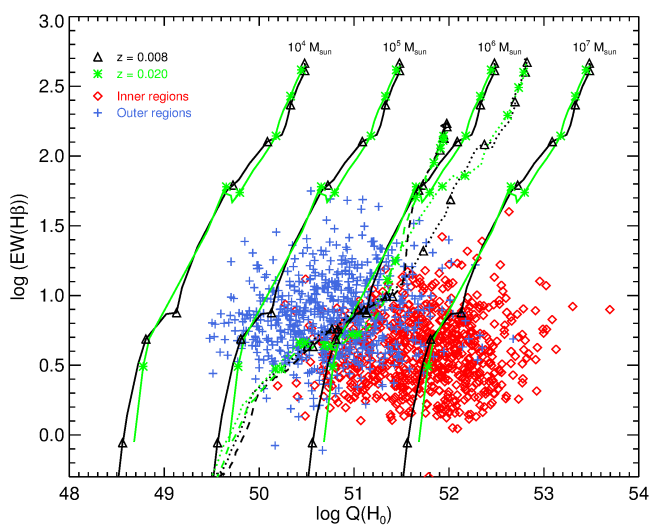}
\caption{Diagram representing EW(H$\beta$) vs $Q(H_0)$ for inner (red diamonds) and outer (blue crosses) observed \hh regions. Overplotted solid lines are predictions by PopStar models for the evolution of EW(H$\beta$) and $Q(H_0)$ depending on age for two different metallicities: \textit{Z} = 0.008 (black triangles) and \textit{Z} = 0.02 (green asterisks) and four different ionising cluster masses, indicated in the plot in units of solar masses. Following the model prediction from top to down, each symbol indicates an evolution of 1 Myr, with a starting age of 1 Myr. Starburst99 model predictions are overplotted for the same metallicities using the same colour code, with dotted lines for M$_{up}$ = 100 M$_{\odot}$ and dashed lines for M$_{up}$ = 30 M$_{\odot}$.}
\label{Fig: EW(Hb) vs Q(H)}
\end{figure}

The strong degeneracy between age and ionisation parameter and the influence of the WR prevents us to estimate ages for the studied regions with the only use of the strong emission-line ratios employed in the diagnostic diagrams. However, the H$\alpha$ and H$\beta$ equivalent widths (EW(H$\alpha$) and EW(H$\beta$)) are good age indicators for ionised regions, as they decrease with the evolution of the ionising cluster, thus providing useful information to study possible differences in the evolutionary stages of inner and outer region samples \citep[see e.g.][]{1981Ap&SS..80..267D}. 

As explained in Sect. \ref{Subsec: Ionization parameter}, the number of ionising photons emitted by an ionising cluster is an increasing function of the cluster mass and a decreasing function of the cluster age. The relation between the EW(H$\beta$) and $Q(H_0)$ of our inner and outer regions is represented in Fig. \ref{Fig: EW(Hb) vs Q(H)}. The equivalent widths estimation is described in Paper I and references therein, while the number of ionising photons calculation is described in Paper I and Sec. \ref{Sec: Derived quantities of the observed regions samples}. In order to help interpreting the results, we include the predictions of the PopStar models for two metallicities, \textit{Z} = 0.008 and \textit{Z} = 0.02, that are the ones that encompass the range of oxygen abundances observed in our samples. Model predictions are scaled for four different ionising cluster masses, constraining the range estimated for our regions (see Paper I), and evolution is represented for ages between 1 and 10 Myr, which the estimated duration of \hh regions.

We observe a strong difference between the observed equivalent widths and the predictions made by the models for simple ionising populations during the first five million years of evolution. This may be caused by the presence of non-ionising underlying stellar populations in the observed \hh regions, but several other factors may have an influence that should also be taken into account. In the first place, we are considering models with a Salpeter IMF with M$_{up}$ = 100 M$_{\odot}$. A lower value of this upper limit would reduce the number of massive stars in the cluster, and therefore decrease the number of ionising photons emitted during the first stages of evolution. This would reduce the Balmer emission line intensities, while leaving the continuum rather unaffected, consequently decreasing the line equivalent width of the model predictions. In order to evaluate this effect, we also compare our results with predictions made by the Starburst99 evolutionary population models \citep{1999ApJS..123....3L} for a cluster of 10$^6$ M$_{\odot}$, considering two upper limits of the Salpeter IMF: M$_{up}$ = 100 M$_{\odot}$ and M$_{up}$ = 30 M$_{\odot}$. These models are also plotted in Fig. \ref{Fig: EW(Hb) vs Q(H)}. There is a good agreement between the predictions made by PopStar and Starburst99 models with the same upper limit of the IMF, for both metallicities and the same cluster mass. In the case of the models with the lower M$_{up}$ in the IMF, Starburst99 EW(H$\beta$) predictions are half order of magnitude smaller for the first Myrs of the cluster evolution. When we reach later stages of the evolution, predictions made by the Starburst99 models for the two upper limits of the IMF reunite. 

Other possible factors affecting the values of the predicted equivalent widths are that we are considering no photon escape and no dust absorption. Some studies on particular \hh regions belonging to the Large and Small Magallanic Clouds find that up to 45-50 \% of the ionising radiation escapes the regions, providing an important source to ionise the diffuse interstellar medium of these galaxies \citep{1997MNRAS.291..827O,2002ApJ...564..704R}. Working also with \hh region observations by the CALIFA survey, \cite{2016A&A...594A..37M} compared observed EW(H$\alpha$) with model predictions and interpreted the differences as the result of a median leaking of 80\% of the photons. On the other hand, a significant fraction of the ionising photons is expected to be absorbed by dust particles in the nebula, and some works consider for this fraction a useful mean value of 0.3 \citep{1987A&A...176....1B,1991A&AS...88..399M}. Nevertheless, PopStar models do not account for the effects of dust. All these considerations would imply that an important part of the emitted ionising photons predicted by the models for simple ionising populations are not contributing to the photoionisation of the gas, and could partially explain the discrepancies between predicted and observed equivalent widths.

It can be observed that inner regions have higher numbers of ionising photons, as already seen in Paper I. The dependance of $Q(H_0)$ with the age and the mentioned uncertainties entail that we can only establish lower (see Paper I) and upper (derived from Fig. \ref{Fig: EW(Hb) vs Q(H)}) limits to the ionising cluster masses. Even considering the associated uncertainties, we interpret that inner regions show remarkably higher ionising cluster masses than outer regions. The upper limits of the ionising cluster masses estimated from Fig. \ref{Fig: EW(Hb) vs Q(H)} for the inner regions reach values up to $\sim$10$^7$ M$_{\odot}$, a result that points to the possibility that we may be observing star-forming complexes, instead of individual inner \hh regions, a situation previously observed in the literature \citep{2007MNRAS.378..163H,2009MNRAS.396.2295H}. On the other hand, a certain amount of the outer \hh regions have lower limits of the ionising masses below $\sim$10$^4$ M$_{\odot}$, and clusters with such low mass values may suffer stochastic effects, which would affect the number of massive stars and the hardness of the ionising radiation, thus modifying some emission-line ratios. 

\begin{figure}
\centering
\includegraphics[scale=0.48]{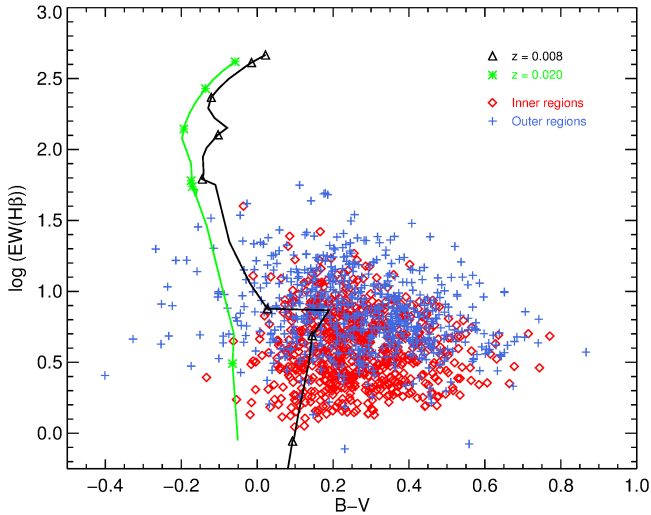}
\caption{Diagram representing EW(H$\beta$) vs B-V colour for inner (red diamonds) and outer (blue crosses) observed \hh regions. Overplotted lines are predictions by PopStar models for the evolution of EW(H$\beta$) and B-V depending on age for two different metallicities: \textit{Z} = 0.008 (black triangles) and \textit{Z} = 0.02 (green asterisks). Following the model prediction from top to down, each symbol indicates an evolution of 1 Myr, with a starting age of 1 Myr. }
\label{Fig: EW(Hb) vs B-V}
\end{figure}

Fig. \ref{Fig: EW(Hb) vs Q(H)} also shows that outer regions have slightly larger values of the H$\beta$ equivalent widths, with a mean value of log EW(H$\beta$) = $0.59\pm0.04$ in the case of the inner regions and log EW(H$\beta$) = $0.83\pm0.06$ for the outers. This result points to younger ages and previous evolutionary stages for the outer regions. Nevertheless, both inner and outer regions show very low values of EW(H$\beta$). This can be caused by the influence of the regions non-ionising populations and/or the galactic underlying populations, which would contribute to the spectra continuum but not to the emission line intensities, thus decreasing the observed EW(H$\beta$). In order to analyse the possible presence of these underlying stellar populations, Fig. \ref{Fig: EW(Hb) vs B-V} shows the distribution of inner and outer observed \hh regions in a diagram of EW(H$\beta$) versus the B-V colour. Colour estimations for both region samples are explained in Paper I. PopStar model predictions are also plotted, following the same scheme employed in Fig. \ref{Fig: EW(Hb) vs Q(H)}. Equivalent widths decrease fast with age for simple ionising stellar populations, as seen by the evolution of the models predictions. On the contrary, colour evolves slowly with age, and the presence of underlying non-ionising populations increases the B-V colour, due to the influence of A stars in the main sequence. The distribution of our observed samples shows that, while outer regions have larger equivalent widths and therefore younger ages, both samples show higher values of the B-V colour than those predicted by models, suggesting the presence of composite stellar populations (ionising and non-ionising). 

\subsection{Ionising and non-ionising continuum contributions}
\label{Subsec: Continuum contributions}

The existence of composite populations implies that the region continua are composed by the sum of an ionising and a non-ionising population contributions. A possible approach to estimate the different contributions of ionising and non-ionising populations  to the observed  continuum is to consider the following:

\begin{equation}
{\rm EW(H\alpha)_{obs}}= \frac{\rm L(H\alpha)}{\rm {I_{cont}(H\alpha)}} = \frac{\rm L(H\alpha)}{\rm {I_{c, ion}(H\alpha)} + \rm {I_{c, non\,ion}(H\alpha)}} =
\end{equation}
\begin{equation}
= \frac{\rm L(H\alpha)/I_{c, ion}(H\alpha)}{1 + (\rm {I_{c, non\,ion}(H\alpha)}/\rm {I_{c, ion}(H\alpha)})} = \frac{\rm EW(H\alpha)_{ion}}{1 + (\rm {I_{c, non\,ion}(H\alpha)}/\rm {I_{c, ion}(H\alpha)})}.
\end{equation}
\\
Therefore the ratio between ionising and non ionising H$\alpha$ continuum intensities can be expressed:

\begin{equation}
\frac{\rm I_{c, non\,ion}(H\alpha)}{\rm I_{c, ion}(H\alpha)}= \frac{\rm EW_{ion}(H\alpha)}{\rm EW_{obs}(H\alpha)}-1.
\end{equation}

\begin{figure}
\centering
\includegraphics[scale=0.48]{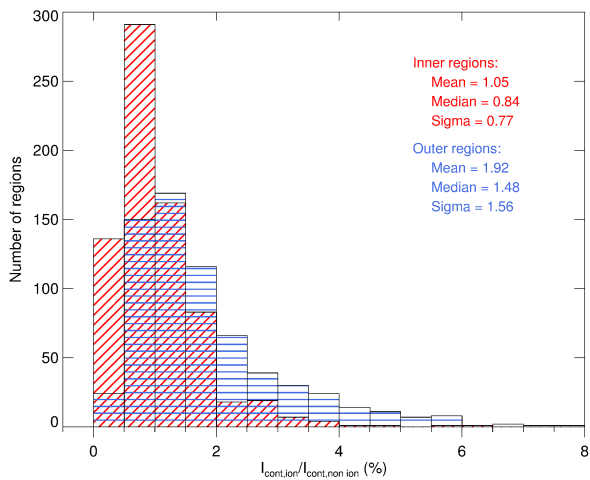}
\caption{Histogram of the percentage of ionising population in relation to non-ionising population found for the inner (red diagonally-hatched diagram) and outer (blue horizontally-hatched diagram) observed region samples, considering model predictions for \textit{Z} = 0.008 and log t = 6.00 (see explanation in the text).}
\label{Fig: Hist continuum components ratio}
\end{figure}

Where $\rm EW_{obs}(H\alpha)$ is the observed equivalent width, and $\rm EW_{ion}(H\alpha)$ is the equivalent width that would correspond only to the ionising population. This value can be obtained from the PopStar model predictions, as a function of age and metallicity. For a metallicity of \textit{Z} = 0.008, we calculate the ratio between non-ionising and ionising continua for all the ages in our time range. Fig. \ref{Fig: Hist continuum components ratio} shows this ratio histogram in percentage for an age of 1 Myr, for both inner and outer observed \hh regions. It is observed that the contribution of the ionising population to the continuum intensity is of an order of 1\% on average. These values should be considered as upper limits of the ionising population contributions, as we are considering that the whole observed H$\alpha$ emission is due to the photoionisation by massive stars, without considering contributions by supernova remnants, shocks or diffuse ionised gas. The contribution of the non-ionising population is larger for the inner regions, as expected due to the influence of the galaxy bulge. Regarding ages, the ratio between non-ionising and ionising continua decreases with evolution of the \hh regions as more stars contribute to it, reaching values one order of magnitude lower for ages of 5.2 Myr.

\section{Derived quantities of the observed \hh regions samples}
\label{Sec: Derived quantities of the observed regions samples}

This work is mainly focused on the differential study of the physical properties of the ionising clusters. Nevertheless, the observational and functional parameters obtained, that provide information about the ionising clusters, also allow us to estimate some of the physical characteristics of the inner and outer \hh regions. Following the expressions by \cite{2000MNRAS.318..462D}, the H$\alpha$ luminosity and the number of ionising photons can be calculated from the observed reddening corrected H$\alpha$ flux and the distance to the corresponding galaxy. Using a distance average value for our galaxy sample of 65 Mpc, both quantities could be expressed as:

\begin{equation}
{L(H\alpha)}= 5.06 \times 10^{39} \left(\frac{F(H\alpha)}{10^{-14}}\right) \left(\frac{D}{65}\right)^{2} {\rm erg\,s^{-1}},
\end{equation}

\begin{equation}
{Q(H_{0})}= 3.70 \times 10^{51} \left(\frac{F(H\alpha)}{10^{-14}}\right) \left(\frac{D}{65}\right)^{2} {\rm photons\,s^{-1}},
\end{equation}

where F(H$\alpha$) is expressed in erg s$^{-1}$ cm$^{-2}$ and D in Mpc. Using these expressions, values obtained for H$\alpha$ luminosities and numbers of ionising photons are coincident with those already represented in Paper I and used in previous sections of this work.

It is also possible to obtain the angular sizes of the \hh regions using the observed reddening corrected H$\alpha$ flux, the ionisation parameter and the electron density:
\begin{equation}
\left(\frac{\Phi}{10"}\right)= 0.19 \left(\frac{F(H\alpha)}{10^{-14}}\right)^{1/2} \left(\frac{u}{10^{-3}}\right)^{-1/2} \left(\frac{n_{e}}{10}\right)^{-1/2},
\end{equation}
where $\Phi$ is the angular diameter in units of 10 arcsec, which does not depend on the distance to a given galaxy. 

The corresponding filling factors, that is, the volume fraction of an \hh region which is occupied by fully ionised matter, can be derived from the observed reddening corrected H$\alpha$ flux, the galaxy distance, the ionisation parameter and the electron density.  The expression for filling factors, using quantities average values corresponding to our observed samples:
\begin{equation}
{\epsilon}= 0.09 \left(\frac{F(H\alpha)}{10^{-14}}\right)^{-1/2} \left(\frac{D}{65}\right)^{-1} \left(\frac{u}{10^{-3}}\right)^{3/2} \left(\frac{\alpha_{B}(H^{0},T)}{10^{-13}}\right)^{-1} \left(\frac{n_{e}}{10}\right)^{-1/2},
\end{equation}

where $\alpha_{B}(H^{0},T)$ is the recombination coefficient for hydrogen for all levels except the ground level, with a value of $\alpha_{B}(H^{0},T)$ = $2.59 \times 10^{-13}$ cm$^{3}$ s$^{-1}$, corresponding to $T = 10000$K \citep{1989agna.book.....O}. We consider a fixed value of $n_{e}$ = 10 cm$^{-3}$, representative of the low electron density regime (see Sec. \ref{Subsec: Parameters grid} and Paper I). 

Finally, the mass of ionised hydrogen for every region can be estimated using the expression by \cite{1989agna.book.....O} and introducing observed and derived quantities \citep[see][]{2000MNRAS.318..462D}, reaching the expression:
\begin{equation}
{M(\hh)}= 75.14 \times 10^{4} \left(\frac{F(H\alpha)}{10^{-14}}\right) \left(\frac{n_{e}}{10}\right)^{-1} \left(\frac{D}{65}\right)^{2} {\rm M_{\odot}}.
\end{equation}

\begin{figure*}
\centering
\includegraphics[scale=0.5]{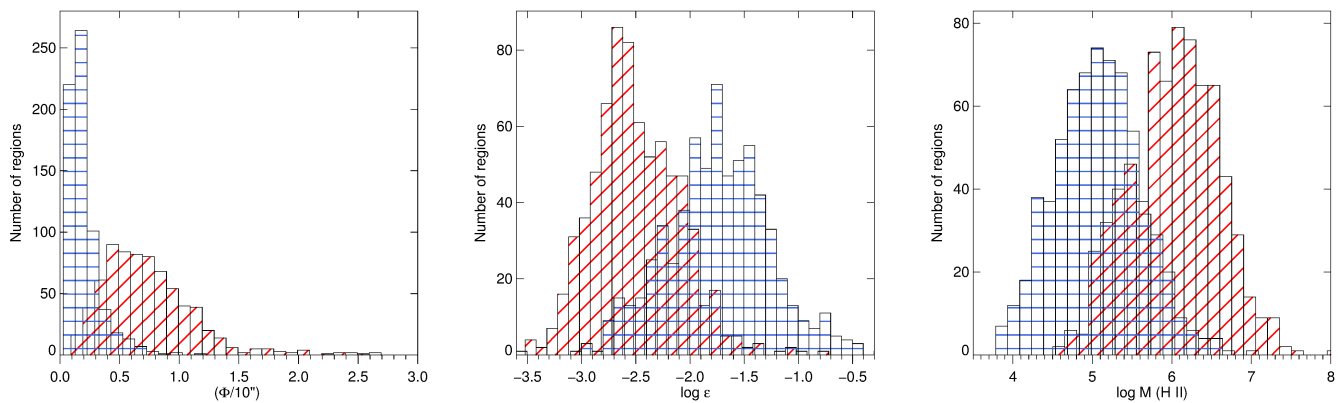}
\caption{Angular sizes (left), filling factors (centre) and mass of ionised hydrogen (right) for the inner (red diagonally-hatched diagram) and outer (blue horizontally-hatched diagram) observed region samples.}
\label{Fig. Histograms physical parameters}
\end{figure*}

Histograms of angular sizes, filling factors and ionised hydrogen masses for inner and outer \hh region samples are represented in Fig. \ref{Fig. Histograms physical parameters}. Regarding sizes, for inner \hh regions the derived angular diameters range from 1 to 20 arcsec, while for outer regions we obtain smaller values, strongly picked at 1 arcsec and reaching in a few cases values as large as 10 arcsec. CALIFA spatial resolution (FWHM) is $\sim3"$, therefore most outer \hh regions would not be spatially resolved. On the contrary, most inner regions are resolved and can reach linear sizes as large as 2.5 kpc, that could in fact correspond to large starforming complexes.

The derived filling factors are low, as is usually the case in extragalactic \hh regions, and turn out to be considerably lower for inner than for outer regions. This points again to the possibility of corresponding to large complexes composed by several regions, difficult to resolve by our segregation procedure in a crowded environment. This would also explain the large H$\alpha$ fluxes found for the inner region population, as well as the corresponding ionised gas masses, that differ in average by more than an order magnitude from the outer region sample.

Due to the high values obtained for the \sii~$\lambda$6717/\sii~$\lambda$6731 for both inner and outer regions (see Paper I), we are assuming n$_{e}$ = 10 cm$^{-3}$ in all our calculations. Nevertheless, slightly higher densities could be considered within the low-density plateau existing in the relation between the \sii~ lines ratio and the electronic density \citep{2006agna.book.....O}, up to a value of n$_{e}$ = $\sim$100 cm$^{-3}$. Assuming this density, the filling factors and the angular sizes would be 0.5 orders of magnitude smaller than those obtained for n$_{e}$ = 10 cm$^{-3}$, and the masses of ionised hydrogen would be one order of magnitude smaller. In the case that inner regions are the ones having larger values of the electronic densities, as could be expected regarding previous results in the literature, this would enlarge the difference between the filling factors  of inner and outer regions and would reduce the difference between angular sizes and masses of ionised hydrogen by the mentioned factors.

\section{Summary and conclusions}
\label{Sec: Conclusions}

This is the second part of a comparative study between two samples of 725 inner and 671 outer \hh regions, belonging to 263 isolated spiral galaxies observed by the CALIFA survey \citep{2012A&A...538A...8S}. The galaxy sample and both \hh region samples are defined and characterised in \cite{2018A&A...609A.102R}.

Using the measured strong emission-line ratios, we have estimated the functional parameters of the observed region samples: oxygen abundance, ionisation parameter and effective temperature, although in this last case only a lower limit could be obtained. These functional parameters allow us to study physical properties of the ionising clusters. In order to do that, we have computed 540 photoionisation models with the photoionisation code Cloudy \citep{2017RMxAA..53..385F}, using the derived functional parameters ranges to define the input parameter grid. We consider as ionising sources the SEDs given by the PopStar evolutionary synthesis models \citep{2009MNRAS.398..451M} with a Salpeter IMF, and compute models for the first 5.2 Myr of the clusters evolution, with metallicities between \textit{Z} = 0.006 and \textit{Z} = 0.018 (solar), ionisation parameters between -3.75 and -2.5 in logarithmic units, and an hydrogen density of 10 cm$^{-3}$. Secondary production of nitrogen is taken into consideration. 

Several diagnostic diagrams are employed for the comparison between observed data and model predictions. Higher metallicities are confirmed for the inner \hh regions, but some important particularities are observed regarding different empirical indicators. 

Firstly, the \nii~$\lambda$6583/H$\alpha$ emission-line ratio seems to trace the metallicity for the outer regions, but the saturation of the \nii~at solar metallicities and beyond may produce an underestimation of the inner regions oxygen abundances, which may affect the range of reliability of oxygen abundance calibrations based on this ratio and prevent their application when inner regions are considered. 

In the second place, the O3N2 index seems to trace metallicity for both inner and outer regions, thus it is confirmed as a better abundance tracer than the \nii/H$\alpha$ ratio. However, the degeneracy introduced in the models by age and ionisation parameter avoids a quantitative estimation of the metallicity, and any empirical calibration based on this index may be influenced by the physical properties of the \hii~ region sample employed for its derivation. 

The \nii~$\lambda$6583/\oii~$\lambda$3727 ratio is also confirmed as an oxygen abundance tracer for both inner and outer regions. Nevertheless, a considerable amount of inner regions show \nii~$\lambda$6583/\oii~$\lambda$3727 values that are higher than those predicted by the models, even in the case of the higher input metallicity. We consider this is a consequence of an enhancement of the N/O relative abundances, and interpret it as a new evidence of a non-linear relation between N/O and O/H, with a steeper relation for the higher oxygen abundances. This effect has already been detected by previous works in the literature, and should be considered when studying circumnuclear and/or high metallicity star-forming regions.

Observational data clearly reproduce the turnover of the \sii~$\lambda$6717,6731/H$\alpha$ emission-line ratio in the diagram versus \oiii~$\lambda$5007/H$\beta$. Outer \hh regions are located on the upper branch and the turnover region, while inner \hh regions are mostly located in the lower branch. Models are able to reproduce the upper branch trend, but most inner regions, at the end of the lower branch, are not fitted by any of the model predictions. This double-valued behaviour of the \sii~$\lambda$6717,6731/H$\alpha$ is also observed in its relation with the oxygen abundance. Ionisation parameter calibrations based on this line ratio and metallicity do not reconstruct this bivalued behaviour, reproducing only the low metallicity branch. Therefore, these calibrations should be considered reliable only for metallicities up to the turnover point, and not to be applied when inner regions are considered. 

Lastly, the relation of the \oii~$\lambda$3727/\oiii~$\lambda$5007 emission-line ratio, tracer of the ionisation parameter, with the O3N2 index reproduces the expected relation of increasing \textit{u} for decreasing \textit{Z}. Nevertheless, inner regions show lower \oii~$\lambda$3727/\oiii~$\lambda$5007 ratio values than expected considering model predictions and the trend drawn by the outer regions. We interpret two possible causes for this behaviour: a change in the relation between ionisation parameter and the \oii/\oiii~ratio for the inner regions physical conditions, or a non-linear relation between ionisation parameter and oxygen abundance along the galactic radius.

The mentioned discrepancies observed between diagnostic diagrams in terms of metallicity and ionisation parameter estimations indicate that the intensive and sometimes exclusive use made in the literature of the \oiii~$\lambda$5007/H$\beta$ versus \nii~$\lambda$6583/H$\alpha$ diagnostic diagram may provide conditioned and partial information. Therefore, the consideration of several diagnostic diagrams involving strong emission-lines reflecting a wider variety of physical effects is recommended.

The age of the ionising cluster cannot be determined through comparison with model predictions in the diagnostic diagrams, due to the strong degeneracy existing between the effects of the age and the ionisation parameter in the model predictions and the influence of the WR, that produce a temporal hardening of the ionising spectra. In order to break this degeneracy and to study possible differences in the evolutionary stages, we represent evolutionary diagrams involving the H$\beta$ equivalent widths. The low EW(H$\beta$) values observed for both inner and outer regions and the comparison with model predictions indicate that both samples have composite populations, with ionising and non-ionising stellar population contributions, a result confirmed by the study of the B-V colours. The influence of the ionising contribution to the underlying continuum is of an order of 1\% to the total, and it is lower in the case of the inner regions, which we consider a consequence of the galaxy bulge contribution. Outer regions show larger equivalent widths, pointing to younger ages, but the existing degeneracy between the effects of evolution and underlying populations prevents a quantitative determination of the regions evolutionary states. 

Even considering the uncertainties due to the impossibility of disentangle the effects of evolution and underlying populations, which only allow the estimation of lower and upper limits, inner regions show larger ionising cluster masses than outer regions. Upper limits obtained for the inner regions indicate that we may be observing clumps of regions, instead of individual ones. In the case of the outer regions the lower limits derived for the masses indicate that some of these outer regions may be affected by stochastic effects, that may influencing their emission spectra.

Using functional parameters and observational magnitudes, we have derived physical properties of the nebulae: angular sizes, filling factors and ionised hydrogen masses. Estimated angular sizes for outer regions are smaller, strongly picked at 1 arcsec and in most cases below CALIFA spatial resolution, which would imply that most of the outer regions are not spatially resolved. On the other hand, inner regions have angular diameters ranging from 1 to 20 arcsec, sizes that could correspond to large star-forming complexes. The obtained filling factors are low, and considerably lower for inner than for outer regions. This result, along with the large ionised gas masses also obtained for the inner regions, point again to the possibility of dealing with large complexes composed by several \hh regions instead of individual \hh inner regions. 

\begin{acknowledgements} We acknowledge financial support for the ESTALLIDOS collaboration by the Spanish Ministerio de Econom\'{i}a y Competitividad (MINECO) under grant AYA2016-79724-C4-1-P. We acknowledge financial support from the Marie Curie FP7-PEOPLE-2013-IRSES scheme, under the SELGIFS collaboration (Study of Emission-Line Galaxies with Integral-Field Spectroscopy).  

\end{acknowledgements}

{}

\end{document}